# Semiconductor channel-mediated photodoping in h-BN encapsulated monolayer MoSe$_2$ phototransistors

*Jorge Quereda\*, Talieh S. Ghiasi, Caspar H. van der Wal and Bart J. van Wees.*

Zernike Institute for Advanced Materials, University of Groningen, Groningen, The Netherlands

\* e-mail: j.quereda.bernabeu@rug.nl

Abstract: In optically excited two-dimensional phototransistors, charge transport is often affected by photodoping effects. Recently, it was shown that such effects are especially strong and persistent for graphene/h-BN heterostructures, and that they can be used to controllably tune the charge neutrality point of graphene. In this work we investigate how this technique can be extended to h-BN encapsulated monolayer MoSe$_2$ phototransistors at room temperature. By exposing the sample to 785 nm laser excitation we can controllably increase the charge carrier density of the MoSe$_2$ channel by $\Delta n \approx 4.45 \times 10^{12}$ cm$^{-2}$, equivalent to applying a back gate voltage of ~ 60 V. We also evaluate the efficiency of photodoping at different illumination wavelengths, finding that it is strongly correlated with the light absorption by the MoSe$_2$ layer, and maximizes for excitation on-resonance with the A exciton absorption. This indicates that the photodoping process involves optical absorption by the MoSe$_2$ channel, in contrast with the mechanism earlier described for graphene/h-BN heterostroctures.

**1. Introduction**

Two-dimensional (2D) transition metal dichalcogenides (TMDs) are very attractive materials for the design of optoelectronic devices at the nanoscale [1–5] due to their optical bandgap spanning the visible spectrum, large photoresponse, and high carrier mobility. The most simple and popular device geometry for 2D TMD phototransistors consists of a monolayer crystal transferred onto a SiO$_2$/Si substrate, with metallic contacts built directly on top of the TMD crystal surface. However, recent works showed that encapsulation of the 2D semiconductor channel between hexagonal boron nitride (h-BN) layers largely improves the electrical performance, optoelectronic response, and device stability [6,7], as it allows to prevent channel degradation due to electrostatic interactions with the metallic electrodes and the SiO$_2$ substrate.[8–10] Thus, h-BN encapsulation is rapidly settling as a new standard for high-quality 2D optoelectronics.

In 2D TMDs the optoelectronic response is often caused by two dominant coexisting effects:[11–15] photoconductivity, where light-induced formation of electron-hole pairs (or charged excitons) leads to an increased charge carrier density and electrical conductivity without changing the Fermi energy $E_F$ of the 2D channel; and photodoping, where the light-induced filling or depleting of charge traps and gap states (present in the surrounding materials and interfaces) causes a shift in $E_F$.[16–19] Since part of the charge-states of traps have very long lifetimes, photodoping typically occurs at longer time scales than photoconductivity.

A recent work showed that, for graphene transistors, using an h-BN substrate instead of the usual SiO$_2$ leads to a large enhancement of photodoping effects [20] due to an exchange of charge carriers between graphene and boron nitride, allowing to optically tune the charge carrier density of graphene.[20,21] Further, in a recent experiment, the photodoping effect was used to tune the Fermi energy in encapsulated MoS$_2$ nanoconstrictions at cryogenic temperatures [19], showing that this effect is not exclusive for graphene/h-BN heterostructures. However, the behavior of this effect at room temperature, as well as its dependence on the illumination wavelength were not yet addressed. Here we investigate the role of photodoping in the optical response of an h-BN encapsulated monolayer (1L) MoSe$_2$ phototransistor at room temperature. Under optical excitation with photon energies above the absorption edge of MoSe$_2$, a large, long-lasting, photodoping effect appears, allowing to increase the MoSe$_2$ charge carrier density by $\Delta n \approx 4.5 \times 10^{12}$ cm$^{-2}$ (observed here as a - 60 V shift of the threshold gate voltage). This effect is especially strong when the device is exposed to light while a negative gate voltage is applied. After turning off the excitation, the device remains photodoped for several days. By testing the dependence of photodoping on the excitation energy, we find that this effect only occurs for excitation wavelengths above the absorption edge of 1L-MoSe$_2$, indicating that the photodoping effect is mediated by optical excitation of this material, in contrast with earlier theoretical descriptions for graphene/h-BN structures [20], where photodoping was attributed to the optical excitation of h-BN impurity states [22–28]. Our results show that long-lasting controllable photodoping can be achieved, even at room temperature, for 2D TMD phototransistors with h-BN substrates, and shed light on the mechanism responsible for this effect.

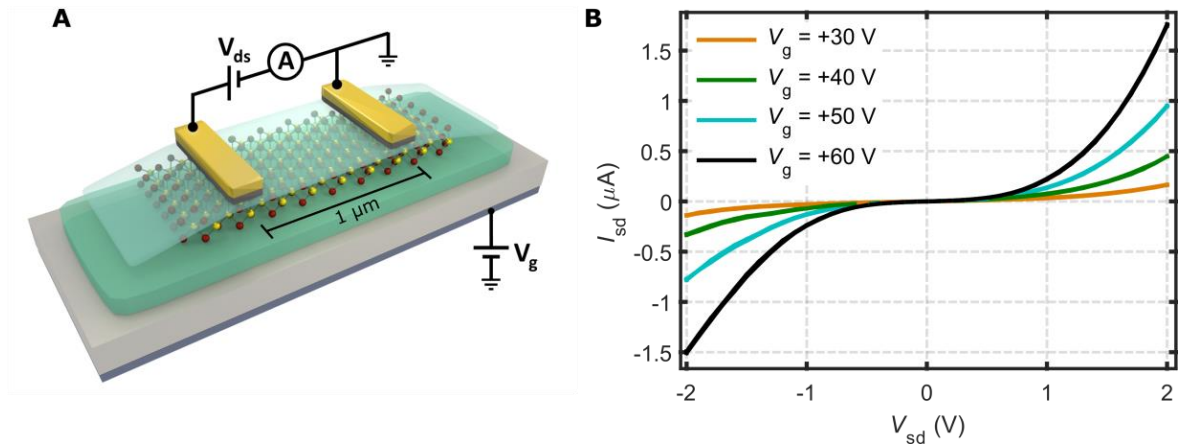

**Figure 1** – (a) Sketch of the h-BN encapsulated monolayer MoSe$_2$ phototransistor. (b) Two-terminal *I-V* characteristics of the device at different gate voltages. The nonlinearity of the *I-V*s is due to the presence of h-BN tunnel barriers at the contacts, as discussed in detail in ref. [20].

## 2. Sample fabrication and electrical characterization

Figure 1a shows a sketch of the studied monolayer MoSe$_2$ phototransistor, where the semiconductor channel is fully encapsulated between bilayer h-BN and bulk h-BN. We used a dry, adhesive-free pick up technique [29] to fabricate the h-BN/MoSe$_2$/h-BN heterostructure on a SiO$_2$ (285 nm)/p-doped Si substrate. Then, we fabricated Ti (5nm)/Au (75 nm) electrodes on top of the structure by e-beam evaporation (see Methods for details). The 5 nm Ti layer allows to achieve a close match between the metal work-function (4.33 eV) and the electron affinity of a 1L-MoSe$_2$ [30]. The thickness of the different layers was characterized by AFM. Figure 1b shows the two-terminal *I-V* characteristic of the 1L-MoSe$_2$ channel at four different gate voltages $V_g$, applied at the bottom Si layer (see Figure 1a), ranging from +30 V to +60 V. The *I-V*s show a non-Ohmic behavior due to the presence of tunnel barriers at the contacts. For a detailed study of the electrical behavior of channel and contacts for this device geometry we address the reader to ref. [7].

## 3. Photodoping and transfer I-V characteristics.

We now investigate the effect of illumination on the transfer characteristic of the 1L-MoSe$_2$ channel. We found that the following procedure is suitable for characterizing the occurrence and persistence of photodoping effect: We ramp the gate voltage from $V_g$ = +70 V to -70 V (trace) and then back to +70 V (retrace) at a ramping speed of 1 V/s while keeping a constant drain-source voltage $V_{ds}$ = 0.5 V and measuring the drain-source current $I_{ds}$. This measurement is first carried out in dark and then repeated upon illumination, as described below. The black curve in Figure 2a shows a transfer characteristic measured while keeping the 1L-MoSe$_2$ device unexposed to light. A clear n-type behavior is observed, with the channel becoming open at a threshold voltage $V_{th}$ = + 25 V (calculated by extrapolating the linear part of the transfer curve and finding its intersection with the horizontal axis, see dashed line in Figure 2a). Very little hysteresis is observed between the trace and retrace measurements, owing to the high quality and environmental stability of the MoSe$_2$ channel. It is worth noting that, at the regions nearby the contacts, $V_{th}$ can increase with respect to the expected value for an unperturbed 1L-MoSe$_2$ channel due to the presence of Schottky and tunnel barriers. Next, we repeat the measurement while keeping the whole device under uniform illumination with a laser power density of 80 pW µm$^{-2}$ and a wavelength λ = 785 nm, matching the A$^0$ exciton resonance of 1L-MoSe$_2$ (red curve in Figure 2a). Under light exposure, photoconductivity is expected to yield a $V_g$-independent increase of the measured current due to the contribution of photogenerated electron-hole pairs, even when the MoSe$_2$ channel is off. Photodoping, on the other hand, changes the MoSe$_2$ Fermi energy, shifting the transfer curve and changing the threshold gate voltage.

While ramping $V_g$ from +70 V to -70 V only a small, constant increase of the drain source current is observed with respect to the transfer characteristic measured in dark, which we attribute to photoconductivity (see inset in Figure 2). Note that for pure photodoping an increase of the off-state current is not expected. When $V_g$ is ramped back from -70 V to 0 V, however, we observe a large shift of the transfer curve towards negative gate voltages due to photodoping. A third transfer curve acquired in dark after exposure to light confirms that the shift persists when the illumination is removed, independently of the $V_g$ ramping direction. As further discussed below, we attribute this shift to a light-induced electron migration from h-BN donor localized states to the MoSe$_2$ valence band.

Next, we characterize the stability of the observed photodoping. Figure 2b shows the time (*t*) evolution of the drain-source current $I_{ds}$ in our device for $V_{ds}$ = 0.5 V and $V_g$ = 0 V after photodoping. The sample is first exposed to illumination at $\lambda = 785$ nm and $V_g = -70$ V for 48 hours and then the light source is turned off and $V_g$ is brought back to 0 V immediately before the measurement starts. As shown in the Figure, $I_{ds}$ decreases over time due to the slow increase of the threshold gate voltage $V_{th}$ (estimated from $I_{ds}$ as discussed below) as photodoping fades away. The time evolution of $I_{ds}$ can be well described by a double exponential function plus an offset:

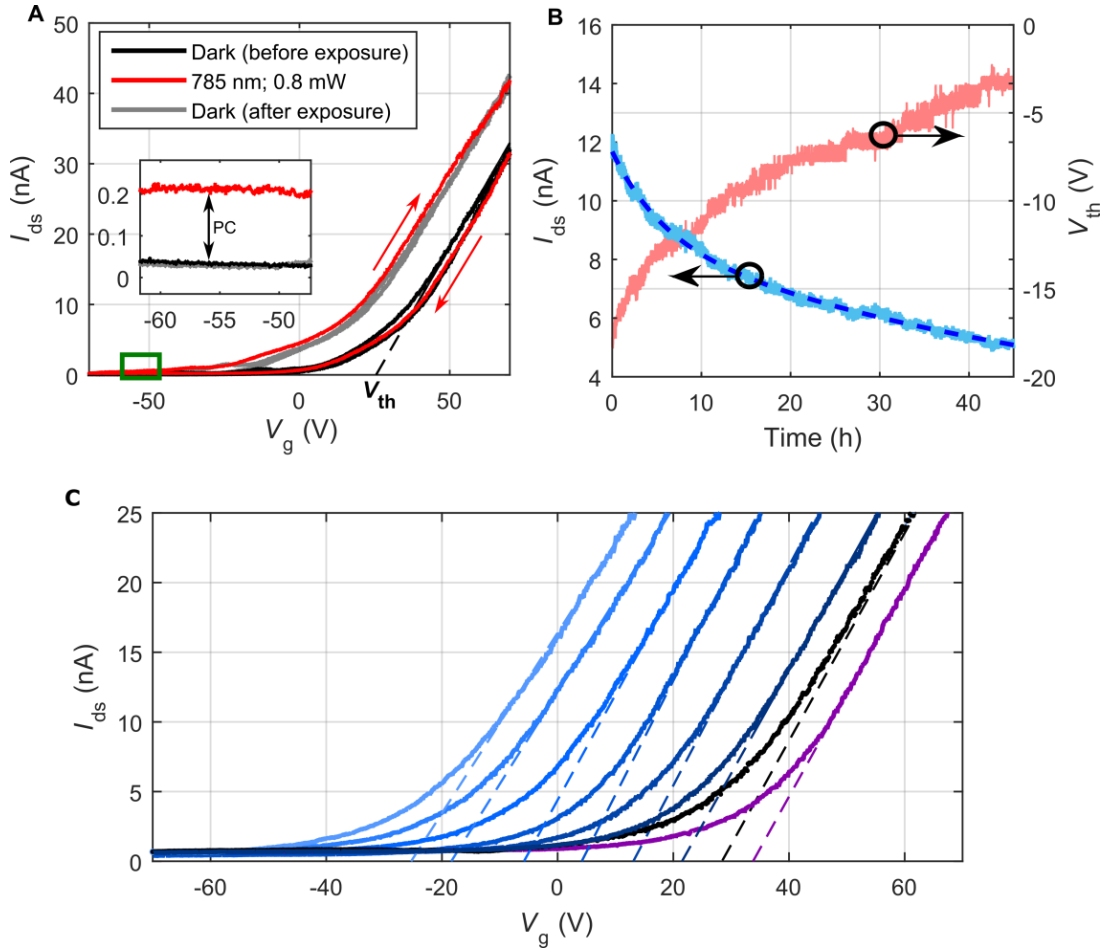

**Figure 2** – Photodoping effect in an h-BN encapsulated 1L-MoSe₂ phototransistor. (a) Transfer curves measured in our device before (black), while (red) and after (gray) exposing the device to light with wavelength $\lambda = 785$ nm and excitation power $P = 0.8$ mW. Inset: Zoom-in of the region marked by a green rectangle in the main panel. (b) Temporal evolution of the drain-source current (blue, left axis) and estimated threshold voltage (pink, right axis) after photodoping, measured in the dark at $V_{ds} = 0.5$ V and $V_g = 0$ V. The dashed blue line is a fit of $I_{ds}$ to a double exponential function (equation 1). (c) Transfer curves measured in dark for different photodoping states. Lighter blue indicates longer time of exposure to light at $V_g = -70$ V (see main text). The black curve corresponds to the device status prior to photodoping. The purple curve is obtained after light exposure at $V_g = +70$ V for 15 hours.

$$I_{ds} = p_1 \exp(-p_2 t) + p_3 \exp(-p_4 t) + p_5 \tag{1}$$

The parameters $p_1...p_5$ are obtained by least square fitting to the experimental data, which yields $p_1 = 7.48$ nA, $p_2 = 0.02$ h$^{-1}$, $p_3 = 3.08$ nA, $p_4 = 0.16$ h$^{-1}$ and $p_5 = 1.10$ nA. The double exponential decay profile of $I_{ds}$ indicates that at least two separate relaxation mechanisms, dominant at different time scales, are involved in this process. We remark that, even at room temperature, the photodoping effect persists for remarkably long times, and even 40 hours after photodoping we get $V_{th} = -3$ V, shifted by 28 V below its value prior to light exposure.

Figure 2c shows transfer characteristics measured in dark after exposing the device to light at $\lambda = 785$ nm and $V_g = -70$ V for different time intervals, between 10 seconds and 2 hours, with the black line corresponding to the state of the device prior to photodoping. As shown in the Figure, the threshold gate voltage $V_{th}$ can be controllably lowered by means of the light exposure time. The largest shift ($\Delta V_{th} = -45$ V) shown in the Figure is reached for an exposure of ~ 2 hours. Similarly, we find that $V_{th}$ can also be increased by exposing the device to light at $V_g = +70$ V. This process allows to recover the original value of $V_{th}$, prior to photodoping after few hours of exposure (shown in Suppl. Info. S1) and even allows to shift the $V_{th}$ slightly above this value. The maximum positive shift ($\Delta V_{th} = +8$ V), corresponding to the purple curve in Fig. 2c, was obtained after 15 hours of exposure at $V_g = +70$ V.

It is worth noting that, apart from the shift of $V_{th}$, the overall shape of the transfer curves remains unmodified for different photodoping states, indicating that the charge carrier mobility is not affected by this process. This allowed to estimate the time evolution of $V_{th}$ from the measured values of $I_{ds}$, as shown in the pink curve of Figure 2b. Further, by using a parallel plate capacitor model, we can estimate the shift in the charge carrier density of MoSe₂ due to photodoping ($\Delta n$). We get

$$\Delta n = -\frac{\Delta V_{th}}{e}\left(\frac{d_{SiO_2}}{\epsilon_0 \epsilon_{SiO_2}} + \frac{d_{BN}}{\epsilon_0 \epsilon_{BN}}\right)^{-1} = 7.42 \times 10^{10} \text{cm}^{-2}\text{V}^{-1} \times \Delta V_{th} \qquad (2)$$

Where $\epsilon_0$ is the vacuum permittivity, $e$ is the electron charge, $\epsilon_{SiO_2} = 3.9$ and $\epsilon_{BN} = 5.06$ are the relative permittivities of SiO₂ and h-BN respectively, and $d_{SiO_2} = 285$ nm and $d_{BN} = 7.5$ nm are the SiO₂ and h-BN thicknesses. Thus, the observed tunability of $V_{th}$ over a range of 60 V corresponds to changing the carrier density by $\Delta n = 4.45 \times 10^{12}$ cm⁻².

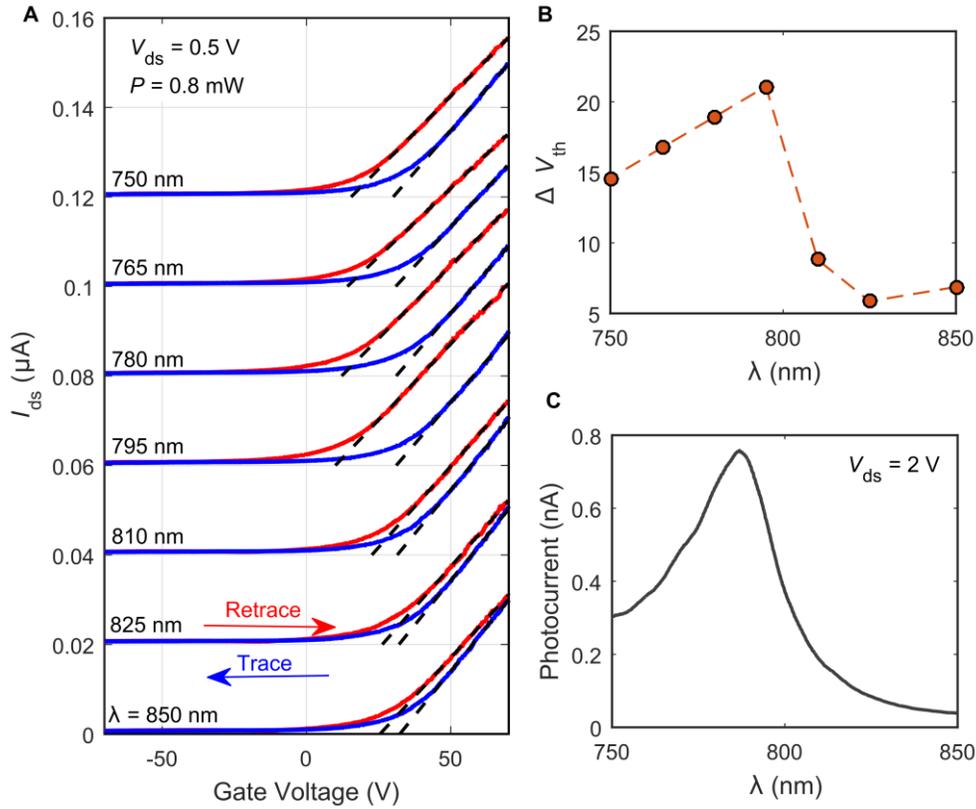

**Figure 3** – Dependence of photodoping on the excitation wavelength. (a) Transfer curves measured in our device under different excitation wavelengths, from 750 to 850 nm. The shift between the trace (blue, from $V_g = +70$ V to $V_g = -70$ V) and retrace (red, from $V_g = -70$ V to $V_g = +70$ V) signals gives an estimation of the effectiveness of photodoping. For all the curves, the system was brought to the same initial photodoping before measuring (b) Shift of the threshold voltage between trace and retrace measurements as a function of the excitation wavelength. (c) Photocurrent spectrum measured in the same device for $V_{ds} = 2$ V and $V_g = 0$ V.

Next, to investigate the spectral response of the observed photodoping we measure transfer characteristics while illuminating the device for a range of excitation wavelengths, from 850 nm to 765 nm (Figure 3a). Before each measurement we keep the system in dark at $V_g$ = +70 V until the same initial threshold voltage $V_{th}$ ≈ 25 V is reached. Then, we ramp the threshold voltage from +70 V to -70 V at a ramping speed of -1 V/s. When the gate voltage is brought to $V_g$ = -70 V while exposing the device to light, the threshold gate voltage $V_{th}$ is lowered due to photodoping (as described earlier and shown in Figure 2a). Figure 3b shows the shift of $V_{th}$ observed between the trace (non-photodoped) and retrace (photodoped) transfer curves as a function of the wavelength, and Figure 3c shows a photocurrent spectrum measured in the same device for comparison (see Methods section and ref. [12] for details). Remarkably, the photodoping-induced shift is strongest for illumination at $\lambda$ = 795 nm (Figure 3b), closely matching the A exciton resonance of 1L-MoSe₂ (Figure 3c), and fades out for wavelengths below the absorption edge of 1L-MoSe₂. This is in marked contrast with the spectral dependence of photodoping reported by L. Ju et al. for graphene/h-BN heterostructures [20], where a detectable photodoping was only observed for illumination wavelengths below 500 nm. In their work, L. Ju et al. attributed the photodoping effect to the optical excitation of electrons from donor-like defects to the h-BN conduction band, followed by a gate-induced migration of these electrons into graphene. Under that description, however, a similar spectral response for photodoping should be expected regardless of whether graphene, MoSe₂ or any other 2D semiconductor is used as channel material, contrary to our observation. Thus, the wavelength dependence observed here strongly indicates that the photodoping process involves the photogeneration of electron-hole pairs in MoSe₂.

In Figure 4, we present an alternative mechanism that allows to account for the observed spectral response. For simplicity, we only consider the bottom h-BN layer, but the same description can be applied without change for the top h-BN layer. First (panel a), the h-BN/MoSe₂ heterostructure is kept in dark and at $V_g = 0$. The Fermi energy $E_F$ of MoSe₂ is set above the neutrality point to account for the n-type doping of the crystal. When a gate voltage $V_g < 0$ is applied (panel b), an electric field appears in the out-of-plane direction, and the Fermi energy of MoSe₂ gets lowered with respect to the edge of its conduction band. Then, if the optical excitation is turned on, with a photon energy above the absorption edge of MoSe₂ (panel c), electron-hole pairs will be formed, either directly or by dissociation of optically generated excitons (see Suppl. Info. S2). If electron-donor localized states are present in the h-BN or at the MoSe₂/h-BN interface (indicated as circles inside the h-BN gap in figure 4), electrons from these states can be transferred to the available states in the MoSe$_2$ valence band. As a result, the h-BN layer will become positively charged. Finally, once the optical excitation is turned off and $V_{th}$ is brought back to 0 V (d), the positively charged h-BN layer will induce a nonzero electric field between the Si and MoSe$_2$ layers, shifting up the Fermi energy of MoSe$_2$ with respect to the conduction band edge. As discussed above and shown in Figure 2b, if the system is kept in dark after photodoping, the h-BN will slowly recover its charge neutrality, as the localized states get filled by charge carriers from the MoSe$_2$ conduction band (indicated by the grey arrow in Figure 4d). However, as experimentally observed, this process is much slower that the electron migration from the localized states to the MoSe$_2$ valence band, yielding a persistent photodoping. The difference in characteristic times for depletion and filling of localized states suggests that these states are more strongly coupled with the MoSe$_2$ valence band than with its conduction band, but the reason for this remains unclear to us. We remark that, as mentioned above, the proposed description is still valid if the h-BN layer is placed on top of the 1L-MoSe$_2$, as in this case a nonzero electric field will still appear in the SiO$_2$ layer as consequence of the light-induced charge accumulation at localized states near the MoSe$_2$/h-BN interface.

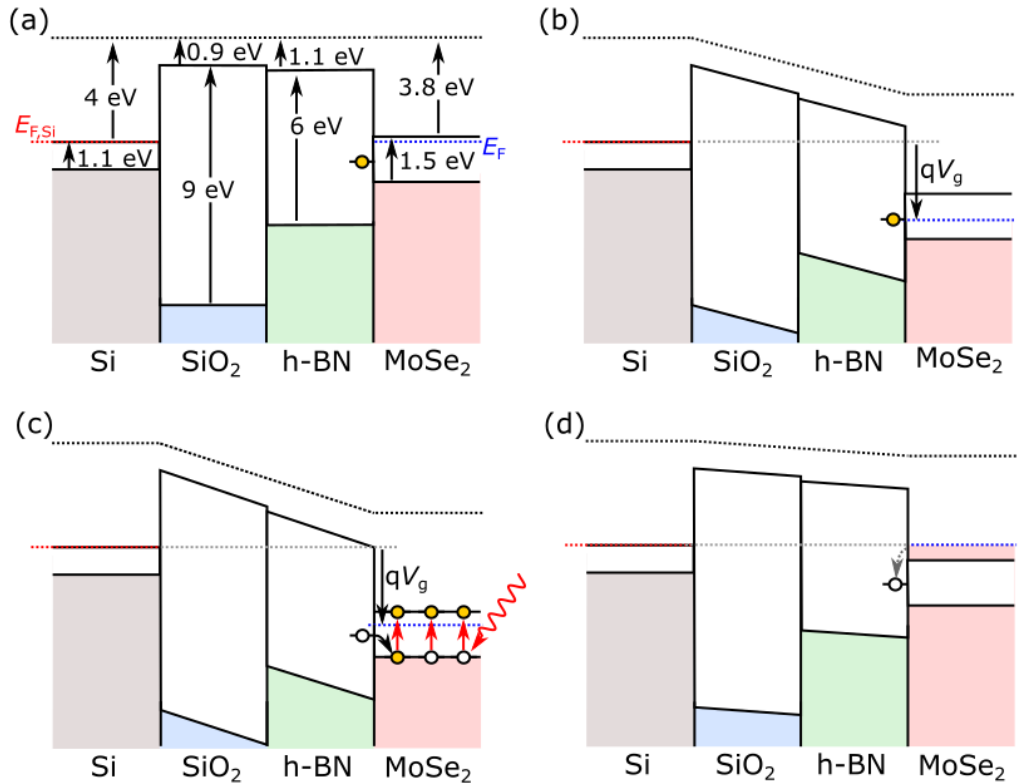

**Figure 4** – Sketch of the proposed photodoping mechanism. Band diagrams of the device (a) in dark, prior to light exposure, with $V_g = 0$ V; (b) in dark, for $V_g < 0$ V; (c) during photodoping with $V_g < 0$ V and $\lambda = 785$ nm; (d) after photodoping, in dark and at $V_g = 0$ V. Yellow (white) circles indicate occupied (empty) electron states.

It is worth noting that a photodoping mechanism similar to the one here described can also appear for SiO$_2$/TMD structures[31]. However, the reported characteristic times for depletion of impurity states in these structures are typically several orders of magnitude lower than those observed here for h-BN substrates. In consequence, photodoping for 2D phototransistors on SiO$_2$ substrates does not yield a persistent shift of $V_{th}$ and manifests instead as a sublinear excitation power dependence of photoconductivity [32]. In our device, we do not expect a measurable contribution to photodoping from SiO$_2$, as this would require the photoexcited charge carriers to migrate from the MoSe$_2$ layer to the SiO$_2$, physically separated by the 7.5 nm thick h-BN layer.

Importantly, in light of the model discussed above, a similar electron migration from h-BN to MoSe$_2$ could also appear in absence of light if some states of the valence band are depleted by applying a sufficiently large negative gate voltage. However, in our devices we do not observe conduction through the valence band (i.e. the gate transfer characteristics do not present ambipolar behavior), indicating that even for the largest applied negative gate $V_g$ = -70 V, the density of depleted states in the valence band is negligible.

## 5. Discussion and final remarks

In all, we demonstrated that photodoping can be used for controllably and persistently tuning the Fermi energy in h-BN encapsulated 1L-MoSe$_2$ phototransistors at room temperature, allowing to tune the carrier density by $\Delta n = 4.5 \times 10^{12}$ cm$^{-2}$. The photoinduced shift of $V_{th}$ was observed up to a few days after exposure to light. The measured spectral response of this effect revealed that photodoping only appears for wavelengths above the absorption edge of MoSe$_2$, clearly indicating that this effect is mediated by optical excitation of the 1L-MoSe$_2$ channel followed by charge migration from h-BN to the channel. Thus, the efficiency of photodoping maximizes for excitation wavelengths at which the 2D channel is highly absorbing.

For the 1L-MoSe$_2$ region directly below the electrodes the optical absorption is expected to be highly suppressed due to screening of the light electric field by the metallic layer. However, should photodoping still occur at these regions, it would produce an increased built-in voltage [7] for the contact, modifying the band alignment between the metallic electrode and the semiconductor channel. This would be observed as an additional contribution to the shift of $V_{th}$ similar to that of the photodoping mechanism discussed above. Note that for this case the migration of electrons from h-BN to the MoSe$_2$ channel is still needed.

It is worth noting that the mechanism proposed here does not require to make any assumption on the nature of the electron-donor localized states, which could be attributed to interfacial states, structural defects or impurities present at the h-BN [22–28,33].

In all, our results demonstrate that the use of h-BN substrates for enhancing the photodoping effect can be easily extended to several two-dimensional semiconductors beyond graphene, and that photodoping is expected to appear for any wavelength at which a significant photoconductivity can occur at the 2D channel. The ubiquity of this effect can also have a negative side, as it could lead to undesired and unexpected optical behaviors. Thus, the role of photodoping should be taken into account when selecting h-BN as substrate for 2D optoelectronic devices.

**Methods**

*Device Fabrication* – We mechanically exfoliate MoSe$_2$ and h-BN from bulk crystals on a SiO$_2$ (285 nm)/doped Si substrate. Then, monolayer MoSe$_2$ and bilayer h-BN are identified by their optical contrast[34] and their thickness is confirmed by Atomic Force Microscopy. We pick up the bilayer h-BN flake using a PC (Poly(Bisphenol A)carbonate) layer attached to a PDMS stamp and later pick up the MoSe$_2$ flake directly with the h-BN surface. Finally, we transfer the whole stack onto a bulk h-BN crystal, exfoliated on a different SiO$_2$/Si substrate. The PC layer is then detached from the PDMS, remaining on top of the 2L-BN / MoSe$_2$ / bulk-BN stack, and must be dissolved using chloroform. For electrode fabrication, we first pattern them by electron-beam lithography using PMMA as resist. Then, we use e-beam evaporation to deposit Ti(5 nm)/Au(75nm) at $10^{-6}$ mbar and lift-off in acetone at 40ºC.

*Photocurrent spectroscopy* –We illuminate the whole sample using a linearly-polarized continuous-wave tunable infrared laser with an illumination power density of 80 pW μm$^{-2}$ while applying a constant drain-source bias, $V_{ds}$ = 1 V. The laser intensity is modulated using a chopper at a frequency of 331 Hz and the light-induced variation of the drain-source current $I_{ds}$ is registered as a function of the illumination wavelength using a Lock-In amplifier.

**Author Contributions**

B.J.v.W. and C.H.v.d.W. initiated the project. J.Q. and T.S.G. had the lead in experimental work and data analysis. The manuscript was written through contributions of all authors. All authors have given approval to the final version of the manuscript.

**Funding Sources**

This research has received funding from the Dutch Foundation for Fundamental Research on Matter (FOM) as a part of the Netherlands Organization for Scientific Research (NWO), FLAG-ERA (15FLAG01- 2), the European Unions Horizon 2020 research and innovation programme under grant agreements No 696656 and 785219 (Graphene Flagship Core 1 and Core 2), NanoNed, the Zernike Institute for Advanced Materials, and the Spinoza Prize awarded to BJ van Wees by NWO.

**Acknowledgments**

We thank Feitze A. van Zwol, Tom Bosma and Jakko de Jong for contributions to the laser control system. We thank H. M. de Roosz, J.G. Holstein, H. Adema and T.J. Schouten for technical assistance.

# Supporting information

Table of contents:

S1. Photodoping at $V_g$ = 70 V

S2. Role of excitons in photoconductivity of MoSe$_2$ phototransistors

**S1. Photodoping at $V_g$ = 70 V**

Figure S1 shows the time evolution of $I_{ds}$ and $V_{th}$ under illumination at $\lambda$ = 785 nm, $V_g$ = 70 V and $V_{ds}$ = 0.5 V. This process allows to accelerate the time for recovery of $V_{th}$ until it reaches its original value, prior to photodoping ($V_{th}$ = 25 V). However, it only allows to increase $V_{th}$ photodoping over this value to a small extent.

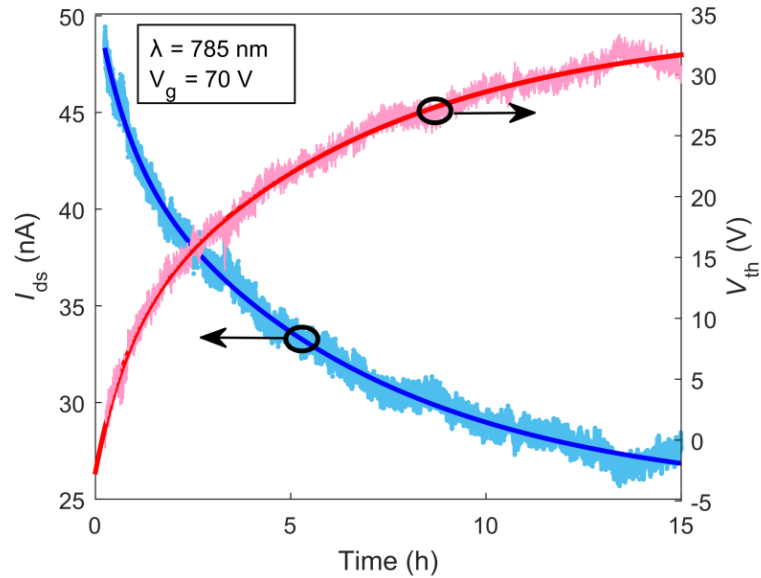

Figure S1 – Time evolution of $I_{ds}$ (blue, left axis) and $V_{th}$ (red, right axis) at 785 nm illumination and $V_g$ = 70 V.

**S2. Role of excitons in photoconductivity of MoSe₂ phototransistors**

As discussed in the main text, photoconductivity relies on the generation of optically excited charge carriers. However, for exciton absorption, photoexcited electrons and holes combine to form excitons with zero net charge. Therefore, in order to contribute to the measured charge current, neutral excitons need to dissociate into free electrons and holes.

It has been proposed that, in monolayer TMD phototransistors, neutral $A^0$ excitons can dissociate due to the large electric gradients formed near the electrodes, giving a nonzero contribution to photoconductivity [S1]. Furthermore, it was shown in prior literature that for TMD/h-BN devices the optical absorption spectrum is mainly dominated by trions, rather than excitons [S2]. Since trions carry a nonzero charge, they can contribute to photoconductivity even without dissociation. Further discussion regarding photoconductivity in 1L-MoSe$_2$ phototransistors can be found in our earlier work [S3].